\documentclass[prb,oneocolumn,showpacs,showkeys,amsmath,amssymb]{revtex4-1}
\usepackage{refstyle}
\usepackage{graphicx}
\usepackage{dcolumn}
\usepackage{bm}
\usepackage{subfigure}
\usepackage{bm}
\usepackage{tabularx}
\usepackage{lettrine}
\usepackage{type1cm}
\usepackage{amsmath}
\usepackage{amsthm}

\begin{document}
\title{Experimental evidence for bulk superconductivity in pure Bismuth single crystal at ambient pressure} 
\author{Om Prakash}
\email[Om Prakash]{op1111shukla@gmail.com}
\address{Department of Condensed Matter Physics and Materials Science,
Tata Institute of Fundamental Research, Mumbai-400005, India}
\author{Anil Kumar, A. Thamizhavel, S. Ramakrishnan}
\address{Department of Condensed Matter Physics and Materials Science,
Tata Institute of Fundamental Research, Mumbai-400005, India}

\begin{abstract}
\textbf{Bulk rhombohedral Bismuth (Bi) at ambient pressure is a semimetal and it remains in the normal state down to 10~mK. The superconductivity (SC) in bulk Bi is thought to be very unlikely due to extremely low carrier density. The question of SC in Bi has remained unsolved both theoretically and experimentally. Here, we report first ever observation of bulk SC in highly pure Bi single crystals (99.9999\%) below 0.53~mK under ambient pressure with an estimated critical magnetic field of 5.2~$\mu$T at 0~K. The conventional Bardeen-Cooper-Schrieffer (BCS) theory cannot explain the observed SC in Bi, since the adiabatic approximation of the BCS theory, $\omega_D/E_F<< 1$, does not hold true for Bi. Bi has a multi-valley type electronic band structure and SC in Bi could be brought about by the inter-valley electron-phonon coupling. Such a scenario calls for new theoretical ideas to understand SC in such low carrier systems with unusual band structure in the non-adiabatic limit, $\omega_D/E_F \gtrsim 1$. The observation of SC in Bi makes it the lowest carrier density superconductor surpassing the record held by doped SrTiO$_3$ for nearly 50 years. }
\end{abstract}
\maketitle

\section*{Introduction}
Bismuth has played a very important role in uncovering many interesting physical properties in condensed matter research \cite{Wilson1932,Mott1936,Edelman1976}, and still continues to draw enormous scientific interests due to its anomalous electronic properties \cite{Li547, Tian2006, Yang1335, Behnia200716, Heremans2000, PhysRevB.58.R10091,Weitzel1991, Muntyanu2006}. Many important phenomena such as Seebeck effect, Nernst effect \cite{Ettingshausen1886}, Shubnikov-de Haas effect, de Haas-van Alphen (dHvA) effect etc. were first discovered in Bi \cite{Fuseya2015}. Determination of the Fermi surface (FS) in Bi using dHvA measurements \cite{Shoenberg1939} provided the basis to determine the Fermi surface of other compounds. The layered structure of Bi plays a crucial role in observing many quantum phenomena rather easily \cite{Wells2009, Li2008, Behnia2007}. Some of the key properties of Bi are: a small density of states (DOS; 4.2$\times 10^{-6}$ states/eV atom) at the Fermi level, very small Fermi surface (FS; $\approx10^{-5}$ of the Brillouin zone, consisting of small electron and hole pockets), low Fermi energy ($E_F \approx 25$~meV), low carrier density ($n \approx 3\times 10^{17}$/cm$^3$ at 4.2~K), and small effective mass for charge carrier ($m_{\rm eff} \approx 10^{-3}$m$_e$, where m$_e$ is the free electron mass) \cite{Liu1995, Smith1964}. Low Fermi-energy in Bi  results in large electronic mean free path (exceeding $2~\mu$m at 300~K) due to the fact that slow electrons are prevented by the conservation laws from interacting with any but the longest lattice vibrations \cite{Sondheimer1952, Pippard1952, Hartman1969}. Moreover, due to small $n$, the Coulomb screening ($\mu^*$) in Bi is much weaker as compared to that of metals, e.g., Au, Cu, Al etc.. 

\begin{figure}[h]
  \includegraphics[width=16cm,angle=0]{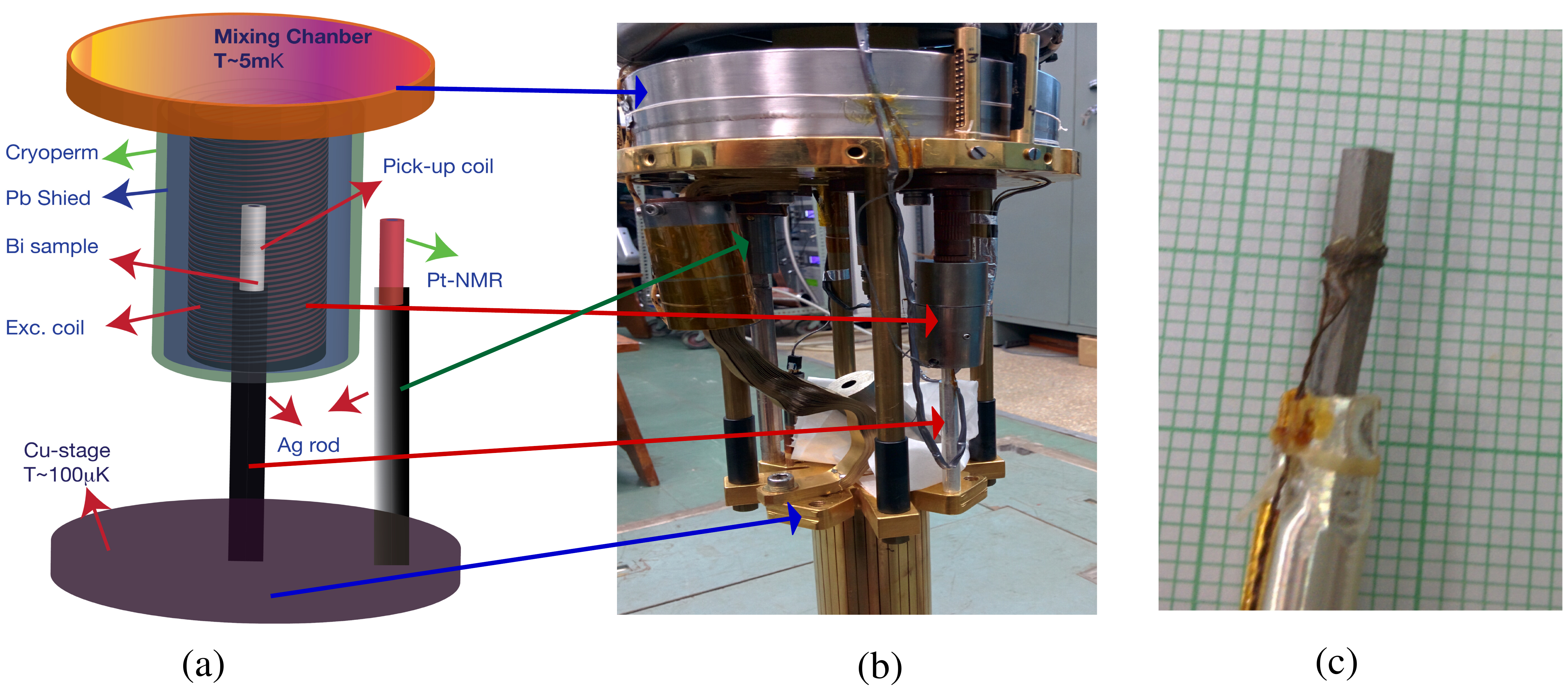}
  \caption{\textbf{Schematic diagram of the magnetic shields and measurement setup:} \textbf{(a)} The measurement setup consists of an excitation coil (0.04$\mu$T/$\mu$A) enclosed in a magnetic shielded environment. The magnetic shielding consists of four layers of Pb foil, inside the two layers of Cryoperm-10 cylindrical shields. This set-up can shield the sample from the external magnetic fields down to 10~nT. The excitation coil and the shielding are attached to the mixing chamber plate of the dilution refrigerator (T$\approx$5~mK). The sample is push-fitted and pinched in a silver rod attached to the low temperature Cu-stage (T$\approx$100$\mu$~K). Similar mounting arrangement has been made for the Pt-NMR thermometer as shown in the diagram. A gradiometer pick-up coil was directly wound on the crystal and connected to the input coil of the dc-SQUID. We used two identical set-ups with different Bi crystals for the measurement.  \textbf{(b)} Actual measurement setup in the dilution insert. The arrows mark different components shown in the drawing.  \textbf{(c)} Bi single crystal attached to the silver (Ag) rod. The  pick-up coil is directly wounded on the crystal and connected to the input terminals of the dc-SQUID.} 
  \label{fig:fig1}
\end{figure}
Search for SC in bulk Bi began more than half a century ago. Although, the SC was observed at high pressures, in amorphous form, thin films, metal hetero-structures, granular nano-wires and nanoparticles of Bi, bulk Bi under ambient conditions remained in the normal state down to 10~mK \cite{Hamada1981,Tian2006, Tian2009, Hakonen1991, Weitzel1991}. Here, we report the observation of bulk SC in pure Bi single crystals ($99.9999\%$) below 0.53~mK by measuring the Meissner effect (diamagnetic) using a gradiometer coil coupled with a dc-SQUID. The Bi single crystals were grown using the Bridgman crystal growth technique and characterized using Energy Dispersive x-ray Spectroscopy (EDX), powder x-ray diffraction (PXRD) and Laue diffraction (see the supplement material) \cite{om2016}. 

For the measurements, Bi crystals of the size $2\times0.2\times 0.2$~cm$^3$ were attached to an annealed high purity silver (Ag) rod (99.999\%) which is threaded to the Copper (Cu) nuclear stage. Rectangular holes of size $0.6\times0.3 \times 0.3$~cm$^3$ were made in the  Ag rods and Bi crystals were push-fitted in the holes along with fine Ag powder for tight sealing. Subsequently, the Ag rod was crimped to hold the samples tight and thus ensuring a good thermal contact. The measurement setup consists of a compensated first order symmetric  gradiometer pick-up coil and an excitation coil, both made up of superconducting niobium (Nb) wires. The gradiometer assembly consists of astatic pair of coils (four turns each) with a distance of 1.2~cm between them. The cross section of the gradiometer coils was minimized to fit the sample and maximize the filling fraction. The gradiometer coils were connected to the input coil of the dc-SQUID (Tristan Technologies, USA) \cite{om2016}. The design and schematic drawing of the measurement setup is shown in Fig~\ref{fig:fig1}(a). The primary coils were wound on the former made from stycast. The whole excitation and pick-up coil assembly was enclosed in magnetic shield consisting of high permeability material called Cryoperm-10 (SEKELS GmbH, Germany) and superconducting lead (Pb) shields. This magnetic shielding arrangement is capable of reducing the external magnetic fields down to less than 10~nT at 4.2~K (see, Fig~\ref{fig:fig1}b), when there is no current in the primary coil. Apart from reducing the effect of external magnetic field the magnetic shields also affect the field inside due to primary coil. For this reason, we calibrated the primary coils enclosed in the magnetic shielding at 4.2~K using very sensitive Single Axis Magnetometer with low field probe (Bartington Instruments Ltd, England, $\pm1$nT resolution) so as to precisely control the excitation magnetic fields during the measurements. The shielded excitation coil set-up is mounted at the bottom of the mixing chamber plate of the dilution refrigerator (Leiden Cryogenics, Netherlands) (Fig~\ref{fig:fig1}(b)). The pick-up coils are connected to the dc-SQUIDs (Fig~\ref{fig:fig1}(c)), which in turn are connected to the RF-amplifier fixed at the head of the cryostat at room temperature. The RF-head is connected to the squid control unit which directly reads output in volts. The dc-SQUID output has been calibrated at 4.2~K by measuring the diamagnetic signal from classical superconductors, Nb and Pb. One of the main challenges in using this method is to calibrate the SQUID output voltage with respect to the susceptibility/Meissner signal. To resolve this, we used Pb  samples of same dimensions as Bi samples and measured the jump in the SQUID output voltage at the transition temperature with different excitation fields. We used the same excitation and pick-up coil set-ups consisting of the magnetic shield mentioned above for the calibration. 

The following two requirements have to be fulfilled in order to observe SC in extremely low $T_C$ superconductors: (1) The sample environment has to be very well shielded from the external magnetic fields as the extremely low $T_C$ superconductors inevitably have very small critical field. Any background magnetic field in the vicinity of the sample can easily suppress the superconducting transition temperature to even lower temperatures; (2) The sample has to be free from  magnetic impurities, since presence of magnetic impurities can also suppress the superconductivity. Apart from these two requirements, the Bi samples need to be extremely pure with no doping. Doping in Bi increases DOS at the Fermi level and can induce superconductivity \cite{Uher1978}. The values of the Hall coefficient $R_H$, measured using the PPMS (Quantum Design, USA) for our single crystals are found to be 0.5~cm$^3$/C (H=0.1T) at 300~K and 3.5~cm$^3$/C (H=0.1T) at 4.2~K \cite{om2016}, which are in agreement with the values reported in literature \cite{Michenaud1972}, suggesting the absence of any doping in our crystals. The estimated Sommerfeld constant, $\gamma \approx 5~\mu$J/mol K$^2$ at 100~mK, for Bi using heat capacity measurements agrees well with the previously reported values \cite{Norman1960} and reflects the high purity of the Bi crystal. The resistivity of the Bi crystals at 300~K, $\rho=1.25\pm 0.02$$\mu \Omega$-m, is also in agreement with the previously reported values for undoped Bi. The residual resistivity ratio (RRR) of as grown samples at 4.2~K is RRR $\approx 100$ indicating high quality of the single crystals. After one week of annealing, we observe significant improvement in the RRR value (RRR $\approx 150$, after annealing), but there was no change observed in the transition temperature within our temperature measurement accuracy of $\pm 10\mu$K.

The experiment was done in a dilution refrigerator equipped with Cu-adiabatic demagnetizing stage. The Cu-stage was first cooled by the dilution refrigerator down to 5~mK followed by magnetization of the Cu-nuclear spins by applying a magnetic field of 9~T using a superconducting magnet (Cryogenics, UK). The Cu-stage was thermally connected to the mixing chamber using an Aluminium (Al) superconducting thermal switch to facilitate isothermal magnetization. The application of the 9~T magnetic field heats up the Cu-stage to nearly 40~mK due to the heat of magnetization and we have to wait for nearly 36 hours to cool down the magnetized Cu-stage to 10~mK.  Subsequently, the Al thermal switch is turned off to thermally disconnect the Cu-stage from the mixing chamber by turning off the current in the solenoid enclosing Al switch. A slow adiabatic demagnetization of Cu nuclear spins over a period of 48-hours cools down the Cu-stage to a base temperature of 100~$\mu$K. Slow demagnetization helps in maintaining thermal equilibrium between the samples and Cu-stage as well as with the NMR thermometer. We used $~^{195}\!$Pt-NMR thermometer for the temperature measurements below 10~mK during adiabatic demagnetization. The NMR thermometer is calibrated against the Cerium Magnesium Nitrate (paramagnetic thermometer) and the SQUID based noise thermometer (MAGNICON GmbH, Germany) at 10~mK. The SQUID based noise thermometer can also measure temperatures down to 1mK and is used along with the NMR thermometer below 10~mK. The details of the adiabatic TIFR nuclear refrigerator with temperature measurement and calibration are given in an earlier report \cite{Naren2012}.  

\begin{figure}[h]
  \includegraphics[width=16cm,angle=0]{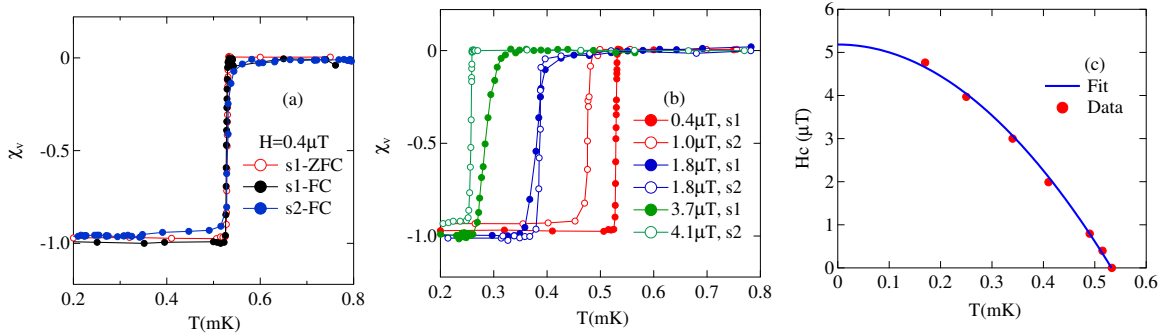}
  \caption{\textbf{Observation of superconductivity, Meissner effect and critical field $H_c(T)$ in Bi single crystals:} \textbf{(a)} The dc-susceptibility ($\chi_v(T)$) of samples s1 and s2 as a function of temperature. A sharp drop in the susceptibility at 0.53~mK marks the transition into superconducting state. \textbf{(b)} The $\chi_v(T)$ as a function of temperature at different magnetic fields. The data corresponding to  $1.8\mu$T magnetic field show the transition at 0.37~mK for both s1 and s2 samples. \textbf{(c)} The magnetic field ($H_c(T)$) vs critical temperature ($T_c$) phase diagram for Bi. The data is fitted to $H_c(T) = H_c(0)[1-(T/T_c)^2$], and the extrapolated critical magnetic field value is $H_c(0)=5.2\pm0.1\mu$T.}
  \label{fig:fig2}
\end{figure}

The superconducting transitions for two Bi samples named s1 and s2 are observed below 0.53~mK in a excitation field of 0.4$\mu$T in the form of sharp drop in the dc-susceptibility $\chi_v$ data as shown in Fig~\ref{fig:fig2}(a). The $\chi_v$(T) data for s1 in the zero field cool (ZFC) and field cool (FC) states suggest the absence of a vortex state, i.e., vortex pinning, indicating type-I superconductivity in Bi. The FC data for s2 is in good agreement with s1 at the same excitation field of 0.4~$\mu$T (Fig~\ref{fig:fig2}(a)). The $\chi_v$(T) for both s1 and s2 was measured in different magnetic fields as shown in Fig~\ref{fig:fig2}(b). The SC transition shifts towards lower temperatures with increasing field. The transition temperatures at different magnetic fields are shown in Fig~\ref{fig:fig2}(c). The data in Fig~\ref{fig:fig2}(c) is fitted to, $H_c(T) = H_c(0)[1-(T/T_c)^2$, to estimate the value of the critical field at 0~K. The critical field value estimated from the fit is $5.2 \pm 0.1 \mu$T. All measurements were performed on the free end of the samples, nearly 1~cm away from the Ag rod \cite{om2016}, avoiding any artefact due to the interface effects. 

We calibrated the measurement set-ups with superconducting Pb and Rh samples of approximately same dimensions. The magnitude of Meissner signal (jump in the SQUID voltage) observed for Bi is nearly the same as the diamagnetic signal observed for superconducting Pb and Rh for the same excitation field of $0.4\mu$T, suggesting that the large volume fraction (bulk) of Bi crystal is undergoing superconducting transition. The extrapolated value of the critical field at 0~K for Bi (H$_C(0)=5.2 \pm 0.1 ~\mu$T) is similar to the critical field of Rh \cite{Buchal1983} even though the Fermi velocity $v_F$, DOS at the Fermi level and carrier density in Bi are very small as compared to Rh. The Fermi velocity of Bi is calculated using the expression, $v_F=(\hbar/m)(3\pi^2n)^{1/3}$, where n is the carrier density. Taking $n=3\times10^{17}/cm^3$, we obtain the value of the Fermi velocity, $v_F= 2.4\times 10^6$ cm/sec for Bi, which is two orders of magnitude smaller than the Fermi velocity in Rh. 

To understand whether SC in Bi is dirty or clean, we estimate the superconducting coherence length using the formula, $\xi_0=\hbar v_F/{3.52k_B T_C}$, assuming the BCS framework. We find $\xi_0=96\mu$m using the value of v$_F$ and T$_C$ for Bi. Since, the mean-free path of Bi as estimated from the resistivity measurements is $\approx 300\mu$m at 4.2~K, the SC transition observed in Bi can be classified as clean type-I superconductor. The BCS model also gives a relation, $B{_C}(0)/T{_C} = (\mu_{0}\gamma/2V{_M})^{1/2}$, where $\gamma$ is the electronic specific-heat coefficient in the normal state and $V{_M}$ is the molar volume. Using the normal state parameters of Bi, we estimate this ratio to be 0.79~mT/K in contrast to the experimental value of 9.4~mT/K, clearly suggesting the inapplicability of standard BCS theory \cite{Bardeen1957}. We would like to mention that the estimation of $\xi_0$ using BCS formula was done to get an idea about the clean nature of superconductivity and the actual value of $\xi_0$ might be different from the value obtained above.

The SC in metallic elements can be understood by the BCS theory \cite{Bardeen1957} and its extensions, and the transition temperature is given by $T_c=\Theta_D$ exp(-$\frac{1}{N(0)V}$), where $\Theta _D$, $N(0)$ and $V$ are the Debye temperature, electronic DOS at $E_F$ and phonon mediated attractive electron-electron interaction, respectively. The general consensus is that even though electron-phonon interaction will be responsible for SC in Bi, the conventional BCS model cannot be applied in Bi. Bismuth has a multi-valley type band structure and small DOS at the Fermi-level. Studying the importance of the multi-valley band structure in low carrier density systems, e.g., Bi, Cohen showed that the attractive electron-electron interaction arising from the exchange of intravalley and intervalley phonons can be larger than the repulsive Coulomb interaction in many-valley semiconductors and semimetals, and can cause these materials to exhibit superconducting properties \cite{Cohen1964}. 

The Fermi-energy $E_F\approx 25~meV$ is comparable to the phonon energy $\hbar \omega_D \approx 12~meV$ in Bi \cite{Zeiger1992}. The BCS-theory of superconductivity is formulated in the so called adiabatic limit, $\omega_D/E_F<< 1$. This assumption is clearly violated for Bi as $\omega_D/E_F \approx~0.5$. Many other superconductors, e.g., SrTiO$_3$, fullerene (C$_{60}$) compounds, high $T_C$ superconductors and superconducting semiconductors, are known with $E_F\lessapprox \hbar\omega_D$. Several attempts have been made to extend the BCS-theory to account for superconductivity in these systems \cite{Gorkov2016, Lin2014, Pietronero1992, Pietronero1995, Koonce1967, Koonce1969} in the non-adiabatic limit. Some other theories on the mechanism of superconductivity based on purely electronic correlations also exist \cite{Kohn1965, Luttinger1966} but cannot be applied to low carrier density systems like Bi. The estimated transition temperature for Bi is orders of magnitude smaller than the observed $T_C$ of 0.53~mK. The work by Pietronero et.al. \cite{Pietronero1992} showed that in the non-adiabatic limit, $\omega_D/E_F \gtrsim 1$, the Migdal’s theorem breaks down and requires the inclusion of vertex renormalization and higher-order diagrams in the self-consistent gap equation. The non-adiabatic effects produce strong enhancement in $T_C$ with respect to the usual Migdal-Eliashberg theory \cite{Migdal1958, Eliashberg1960}. In particular, Pietronero et. al. \cite{Pietronero1995} generalized the many body theory of SC in a perturbative scheme with respect to the parameter, $\lambda\omega_D/E_F$, where $\lambda$ is the electron-phonon coupling constant, by calculating the vertex correction function and self energy in the non-adiabatic limit. They find that the vertex correction function shows a complex behaviour with respect to the momentum (q) and frequency ($\omega$) of the exchange phonon. Specifically, the vertex corrections are positive for small values of $q$ and can lead to strong enhancement of $T_C$, as compared to the usual BCS-theory. In this case the $T_C$ is given by, $T_C = 1.13\Theta_D e^{-1/{\lambda(1+\lambda)-\mu^*}}$, where $\mu^*$ is the Coulomb screening (in the usual BCS-theory, $T_C = 1.13\Theta_D e^{-1/{\lambda-\mu^*}}$). Using the value of $\mu^*=0.105$ \cite{Mata2016} and the observed $T_C = 0.53$~mK in $T_C = 1.13\Theta_D e^{-1/{\lambda(1+\lambda)-\mu^*}}$, we obtain the value of $\lambda=0.16$, suggesting rather weak electron-phonon coupling in Bi. This value of $\lambda$ is similar but smaller than the value estimated for crystalline Bi in a recent simulation study \cite{Mata2016}. In low carrier density systems having multi-valley electronic structures (as is the case for Bi), Cohen \cite{mcohen1969} showed that  inter-valley electron-phonon interactions contribute significantly to the net attractive electronic potential but these inter-valley scattering are associated with large momentum (q) transfer, unlike the situation of enhanced $T_C$ due to vertex corrections at small $q$, as discussed above. Although, the SC in Bi can be qualitatively explained by Pietronero et. al. \cite{Pietronero1995}, new theoretical inputs are needed to estimate the superconducting parameters in low carrier systems in the non-adiabatic limit, which then can be compared with those measured by experiments.

\section*{SUPPLEMENTARY MATERIALS}
\subsection*{Crystal growth and characterization}
\begin{figure}
\includegraphics[width=16cm, angle=0]{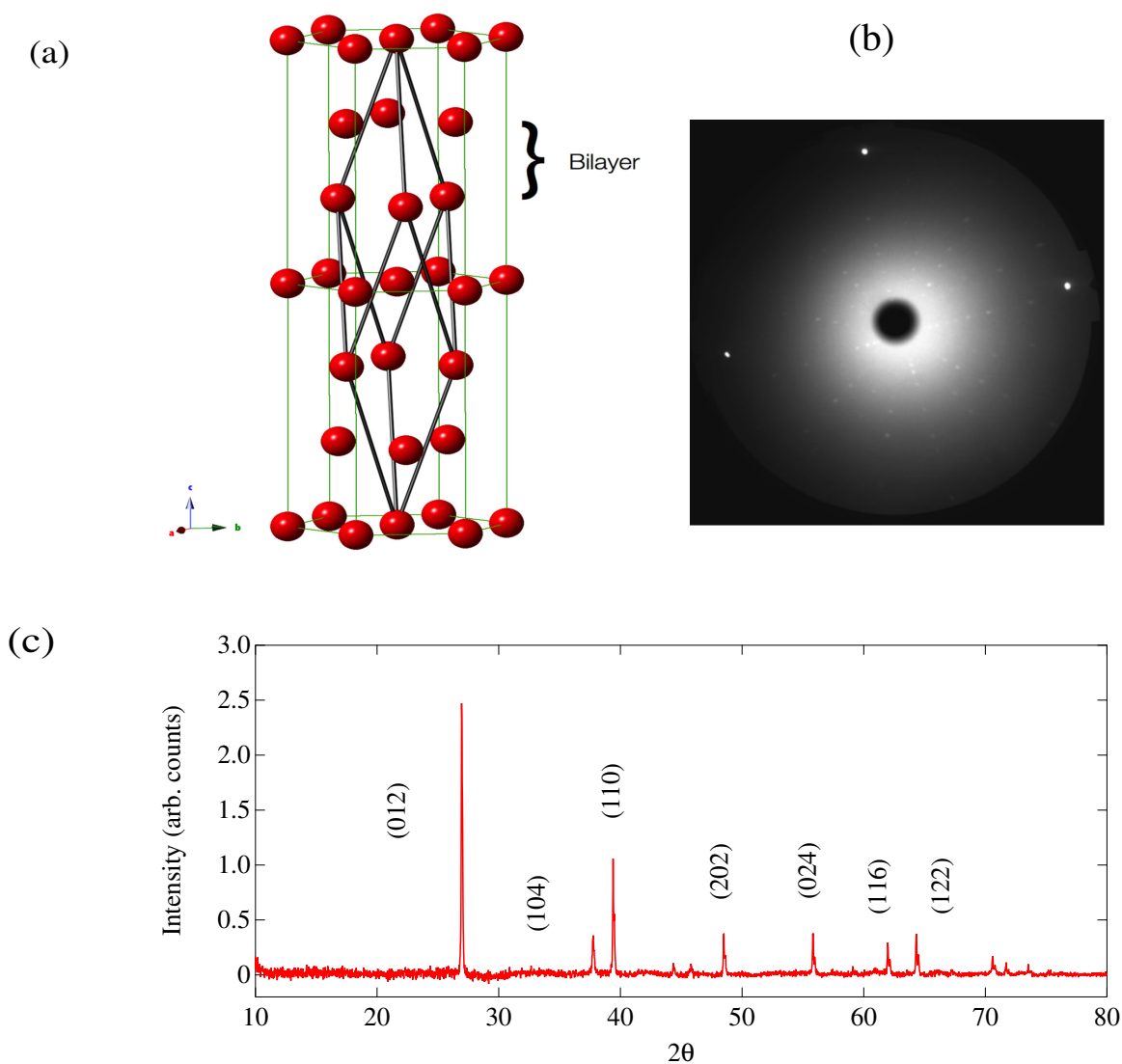}
\caption{\textbf{Crystal structure and characterization of Bi single crystal:} \textbf{(a)} The schematic diagram of hexagonal crystal structure of Bismuth. The rhombohedral unit cell (black) is shown within the crystal structure. Bismuth has two bilayers within the hexagonal structure. \textbf{(b)}  A Laue diffraction pattern of Bismuth crystal for [001] crystallographic direction. The circular spots confirm the single crystalline nature of the grown crystals. \textbf{(c)} X-ray diffraction of Bismuth crystals. All the peaks observed in the recorded pattern can be indexed for the A7 structure and prominent peaks in the xrd pattern are marked.}
\label{fig:fig1S}
\end{figure}

Under ambient condition, Bi is commonly designated as a rhombohedral lattice (space group R-3m, so-called arsenic or A7 structure), which is characterized by a pair of atoms spaced non-equidistantly along the trigonal axis in a Peierls distortion of the simple cubic structure \cite{Fritz2007, Shu2016}. Alternatively, the structure of Bi can be described as a hexagonal lattice with six atoms per unit cell, or as a pseudo-cubic structure with one atom per unit cell\cite{Hofmann2006}. The schematic diagram of the rhombohedral unit cell and hexagonal crystal structure are shown in Fig.~\ref{fig:fig1S}(a). 
\begin{figure}[h]
\includegraphics[width=8cm]{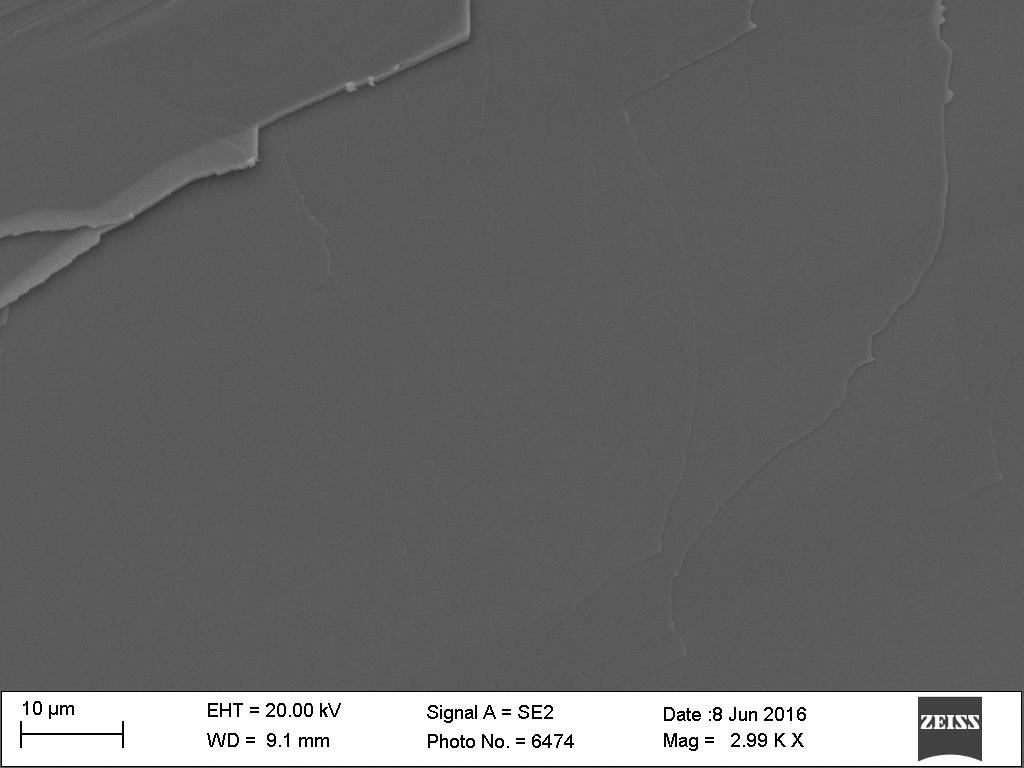}
\includegraphics[width=8cm]{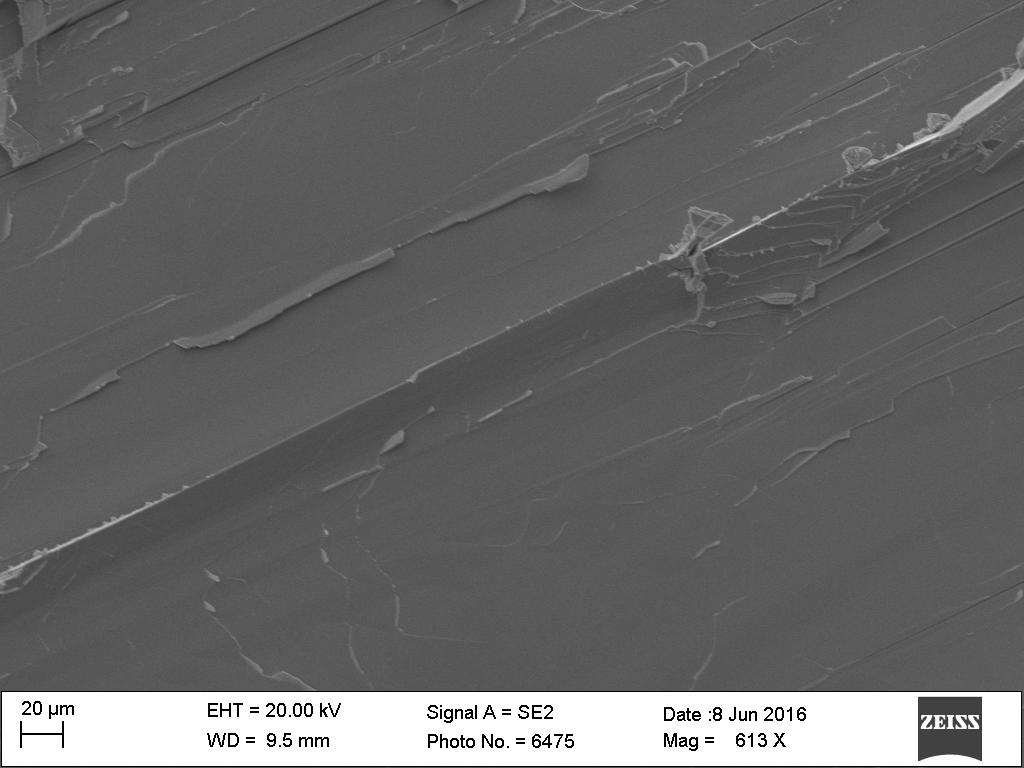}
\includegraphics[width=16cm]{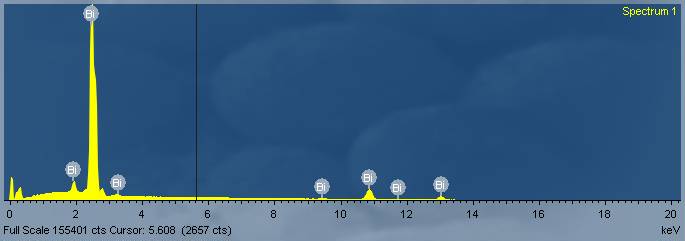}
\caption{\textbf{Sample s1 characterization using EDX:} SEM image of s1-Bi cleaved surface. The EDX spectra was acquired on the front and back surface of the sample. No trace of any elements other than Bismuth was found.}
\label{fig:fig2S}
\end{figure}
\begin{figure}[h]
\includegraphics[width=8cm]{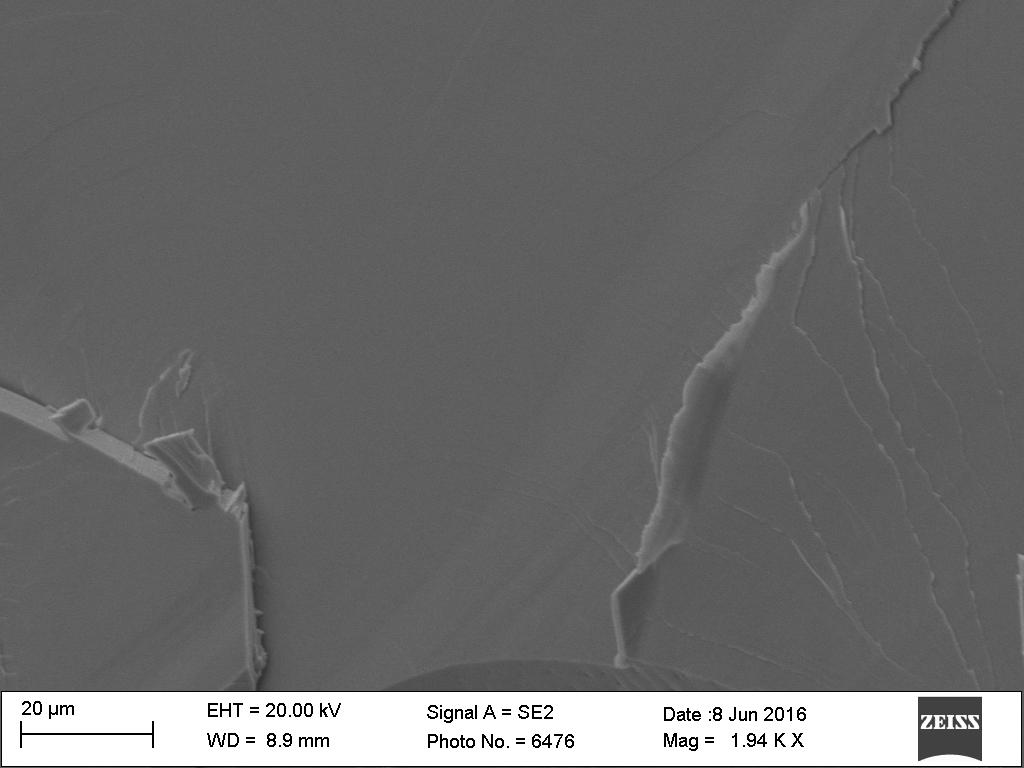}
\includegraphics[width=8cm]{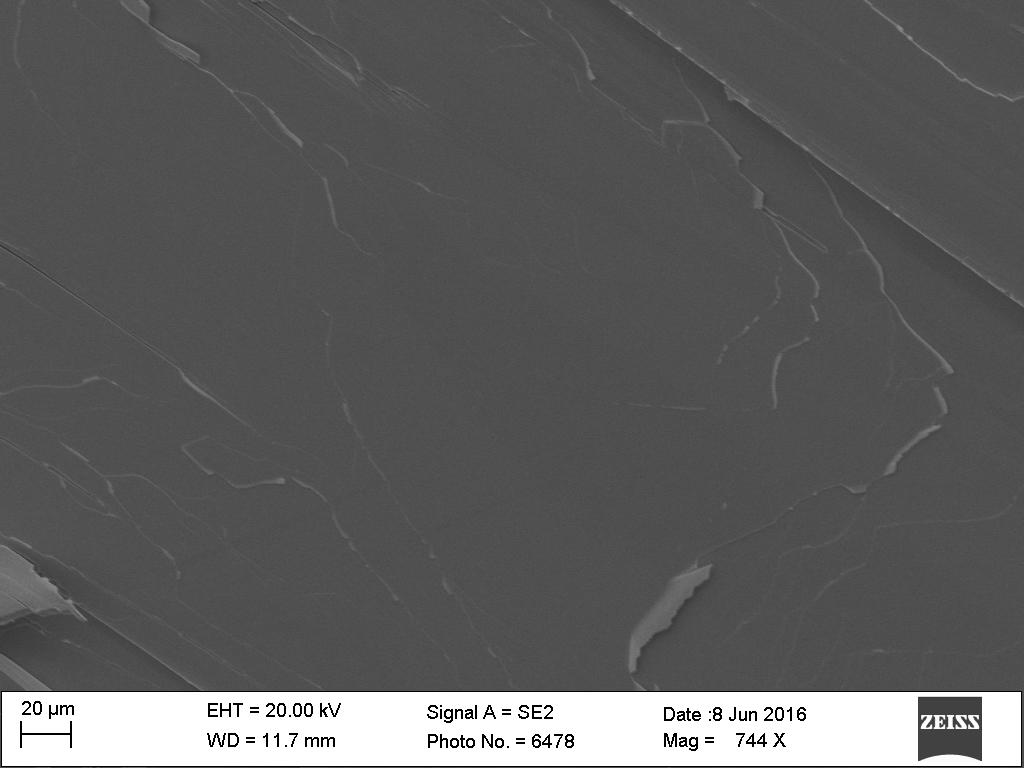}
\caption{\textbf{Sample s2 characterization using EDX:} SEM image of s2-Bi cleaved surface.}
\label{fig:fig3S}
\end{figure}
The Bi single crystals were grown using the Bridgman crystal growth technique. Bi has a low melting point of 271.4$^{\circ}$C and Bridgman technique is suitable for growing Bi single crystals. Highly pure Bi ingots (99.9999$\%$) packed in glass tubes with inert argon were used for crystal growth. Here, we have used quartz tubes with pointed bottom to grow the crystals. Prior to sealing, to avoid oxidation and contamination, the quartz tubes were etched in dilute HF-solution followed by cleaning with distilled water and baked in dynamical vacuum of $5\times 10^{-6}$ mbar, at 1000$^{\circ}$C for 24 hours. The Bi ingots then were transferred from the original sealed tubes to the quartz tubes and the quartz tubes are vacuum ($1\times 10^{-6}$ mbar) sealed. Two such sealed quartz tubes with 2~gm of Bi in each, were kept in a programmable box furnace. Initially, the temperature of the furnace was raised to 600 $^{\circ}$C in 10~h and kept at 600 $^{\circ}$C for 12~h in order to ensure complete melting of Bi. The tubes were then cooled to 350~$^{\circ}$C with the rate of 1$^{\circ}$C/h, followed by cooling to 200$^{\circ}$C at 0.5$^{\circ}$C/h. Slow cooling in the temperature range of the crystallization helps in the getting a large single grain crystal. Subsequently, the furnace was cooled down to 30$^{\circ}$C in next five hours. 

Large crystals of 3-4~mm diameter and 2-3~cm length were obtained. The crystals were stored in a dynamical vacuum of $1\times 10^{-3}$ mbar in desiccators to avoid any contamination. To check the single crystalline nature of the crystals, small portions were cut from both ends to get plane surfaces and exposed to Laue diffraction. The diffraction images obtained from the plane surfaces at both ends of the crystals show patterns corresponding to the [001] crystallographic direction, indicating the present of single grain across the whole length of the crystals as shown in Fig.~\ref{fig:fig1S}(b). The crystals were cut to a rectangular bar of $2\times 0.2\times 0.2$~cm$^3$ using a spark erosion cutting machine. After the cutting all the surfaces were carefully polished and cleaned using ultrasonics to get rid of any surface contamination. The powder x-ray diffraction shows all the diffraction peaks corresponding to the rhombohedral A7 structure \cite{Shu2016} and no extra peaks were observed as shown in Fig.~\ref{fig:fig1}(c). We used cleaved Bi crystals surfaces for characterization using Energy Dispersive x-ray Spectroscopy (EDX). The EDX spectroscopy shows no trace of any impurity elements and secondary phases. The SEM images for both s1 and s2 samples are shown in Fig.~\ref{fig:fig2S} and Fig.~\ref{fig:fig3S} respectively along with the EDX spectra for s1 crystal.

\section*{References}
\bibliography{references}
\bibliographystyle{apsrev4-1}
\section*{Acknowledgements} The authors would like to thank Prof. Sudhansu S. Jha for useful discussions and comments  on the present work. We thank Dr. S. Mukhopadhyay, R. Kulkarni, and D.D. Buddhikot for valuable technical help at the early stages of this work.
\section*{Author contributions}
The project was planned by S.R.. Single crystals were grown and characterized by O.P. and A.T.. All the measurements were done by O.P., A.K.. The data analysis was done by O.P. and S.R.. The manuscript was prepared by S.R. and O.P. and discussed with A.K. and A.T..
\section*{Competing financial interests}
The authors declare no competing financial interests.

 \section*{Correspondence} Correspondence should be addressed to Om Prakash~(email: omprakashshukla@tifr.res.in) and S. Ramakrishnan~(email: ramky@tifr.res.in).
\end{document}